%% file: main.tex
\documentclass[universe,review,accept,pdftex,moreauthors]{Definitions/mdpi} 

\firstpage{1} 
\pubvolume{1}
\issuenum{1}
\articlenumber{0}
\pubyear{2025}
\copyrightyear{2025}
\datereceived{ } 
\daterevised{ } 
\dateaccepted{ } 
\datepublished{ } 
\hreflink{https://doi.org/} 

\Title{The Potential for Hadronic Particle Acceleration in Galactic Pulsar Wind Nebulae}

\TitleCitation{The Potential for  Hadronic Particle Acceleration in Pulsar Wind Nebulae}

\Author{Alison M.W. Mitchell $^{1,\ddagger}$\orcidA{} and Samuel T. Spencer $^{1,2,\dagger,\ddagger}$\orcidB{}}

\AuthorNames{Alison Mitchell and Samuel Spencer}

\isAPAStyle{%
       \AuthorCitation{Mitchell, A.M.W. \& Spencer, S.T.}
         }{%
        \isChicagoStyle{%
        \AuthorCitation{A.M.W. Mitchell and S.T. Spencer.}
        }{
        \AuthorCitation{Mitchell, A.M.W.; Spencer, S.T.}
        }
}

\address{%
$^{1}$ \quad Friedrich-Alexander-Universität Erlangen-Nürnberg, Erlangen Centre for Astroparticle Physics, Nikolaus-Fiebiger-Str. 2, D 91058 Erlangen, Germany\\ 
$^{2}$ \quad Department of Physics, Clarendon Laboratory, Parks Road, Oxford, OX1 3PU, United Kingdom}

\corres{Correspondence: alison.mw.mitchell@fau.de; samuel.spencer@cta-observatory.org} 

\firstnote{Now at Cherenkov Telescope Array Observatory}  
\secondnote{These authors contributed equally to this work.}

\abstract{Pulsar wind nebulae (PWNe), formed when the wind originating from a rapidly rotating neutron star flows out into its surroundings, have now been observed across the electromagnetic spectrum from the radio to the PeV gamma-ray regime. For most of these sources, leptonic processes, where electrons interacting with background photon fields produce high-energy photons through inverse Compton scattering, are believed to be the origin of associated very-high-energy gamma-ray emission. As such, these objects cannot contribute significantly to the galactic hadronic cosmic ray flux at $\sim$ TeV-PeV energies. However, in a handful of cases, the possibility for an energetically sub-dominant hadron population being accelerated and producing very to ultra-high energy gamma-rays through pion decay has not yet been comprehensively excluded. Such scenarios have received renewed attention in the light of recent results from the Large High Altitude Air Shower Observatory (LHAASO). In this review we explore the theoretical background positing hadronic acceleration in galactic PWNe, considering cases where the hadrons escape from the pulsar surface and/or are accelerated in the wind, as well as potential `shock mixing' scenarios. We also explore current and future possible constraints on a hadronic component to PWNe from observations.}


\begin{document}
\include{journalcommands}

\section{Introduction to pulsars and the formation of Pulsar Wind Nebulae}
\label{sec:intro}

Pulsars are rapidly rotating neutron stars with a typical mass of $\sim 1.4 \mathrm{M_{\odot}}$ \citep{universe5070159}, believed to be formed during core-collapse supernovae where the progenitor's stellar mass is in the range $\sim 10-25\,\mathrm{M_\odot}$. Pulsars have strong magnetic fields, ranging from $10^8\,\mathrm{G}$ for millisecond period pulsars (believed to be formed when older pulsars are `spun up' by accretion from a stellar companion) to $10^{15}\mathrm{G}$ for so-called `magnetars' (originally postulated to explain transient `soft-gamma repeater' events) \citep{gaensler_slane_review,10.1093/mnras/275.2.255}. Pulsars typically have currently observed rotation periods $P$ in the range 1.4\,ms \citep{doi:10.1126/science.1123430} to 76\,s \citep{2022NatAs_LongPeriodPSR}, 
but are believed to be formed with initial rotation periods $P_0$ in the range $\sim$10-100\,ms \citep{10.1111/j.1365-2966.2007.12821.x,Spencer_2025}. Pulsars thus possess a significant reservoir of rotational energy, given by
\begin{equation}
E_{\rm rot}=\frac{I\Omega^2}{2},
\end{equation}
where $I$ is the neutron star moment of inertia (typically $\sim 10^{45}\,\mathrm{g\,cm^{-2}}$ \citep{MOLNVIK1985239}) and $\Omega$ is the angular velocity. In the absence of accretion, pulsars' rotation periods increase over time at a rate $\dot{P}$ (typically in the range $\sim10^{-20}-10^{-11}\mathrm{s\,s^{-1}}$ \citep{manchester_atnf_2005,Younes_2017}), primarily due to angular momentum loss via magnetic dipole radiation (which cannot propagate through the PWN or the ISM due to its very low frequency $\nu<1\,\mathrm{kHz}$ \citep{2016era..book.....C}) but also through particle acceleration and potentially gravitational wave emission \citep{Contopoulos_2006}.

\begin{figure}
\centering
\includegraphics[width=0.9\columnwidth]{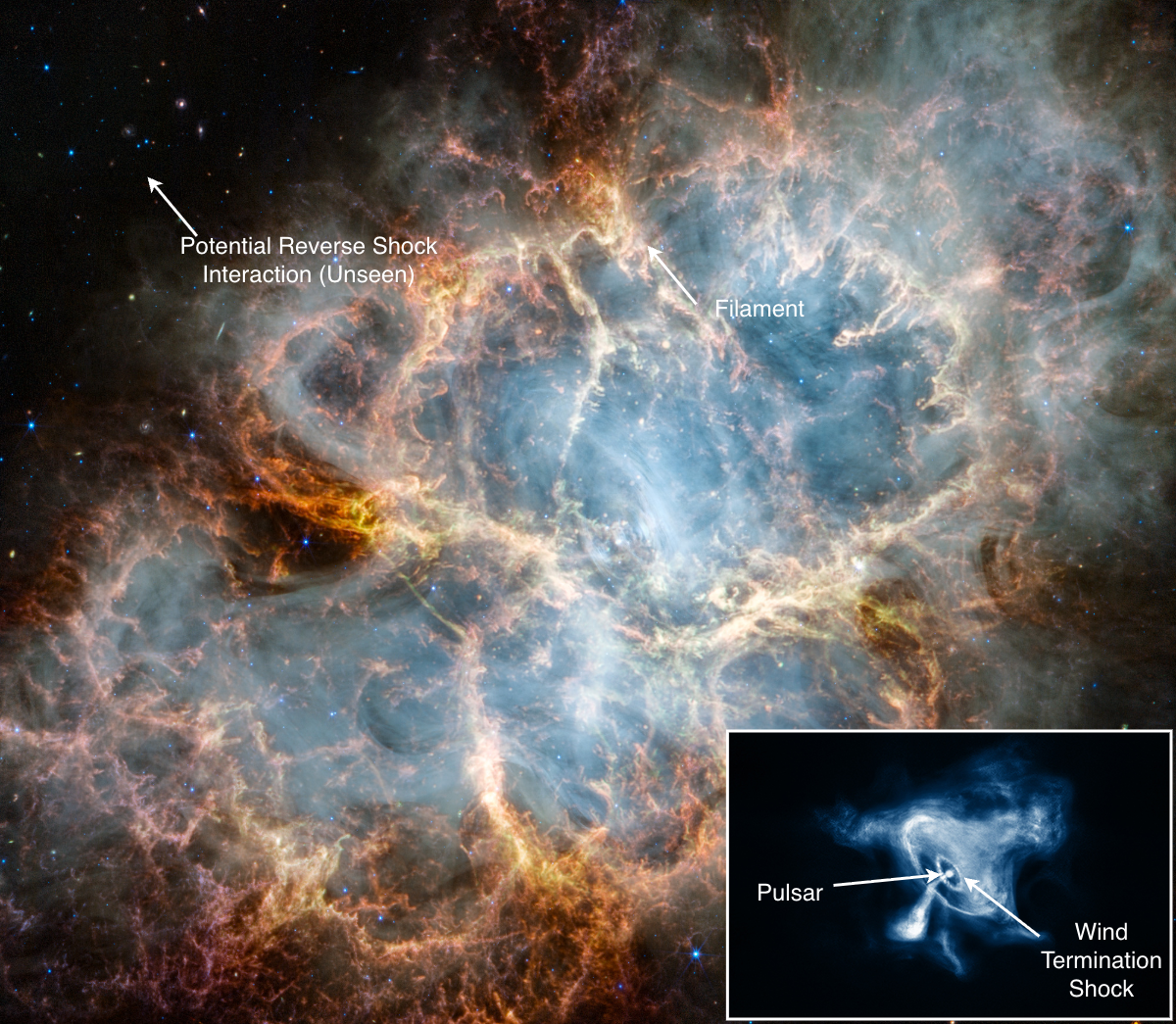}
\caption{The potential acceleration sites in a PWN, as seen for the Crab Nebula by Webb and Chandra. Image credits: Webb: NASA, ESA, CSA, STScI, T. Temim (Princeton University), Chandra: NASA, CXC, SAO. The inset shows solely the X-ray emission. }
\label{fig:struct}
\end{figure}

Pulsars are surrounded by a `light cylinder', defined as the maximum radius at which particles can co-rotate with the pulsar without exceeding the speed of light. Its radius is given by \citep{1969ApJ...157..869G} 
\begin{equation}
r_{\rm lc} \sin\theta = \frac{c}{\Omega},
\end{equation}
where $\theta$ is the polar co-ordinate measured from the rotation axis, $c$ is the speed of light and $\Omega$ is the pulsar's angular velocity. $r_{\rm lc}$ is around 5000\,km for a $P=100\,\mathrm{ms}$ pulsar. The extreme electric field generated by the pulsar is strong enough that it can strip electrons, protons and potentially even ions from the metal-rich pulsar surface \citep{1969ApJ...157..869G, 2015JCAP...08..026K}, however heavier ions can be photodissociated by an X-ray field surrounding the pulsar if it has a sufficiently high temperature \citep{bp97}. The light cylinder is therefore expected to be filled with plasma, where the particles are electromagnetically accelerated and can then escape along open magnetic field lines forming the ultra-relativistic \textit{pulsar wind} \citep{1969ApJ...157..869G}. The pulsed radio emission from the pulsar is thought to originate from curvature radiation from accelerated electrons and positrons a relatively short distance (approximately 1 neutron star radius, $\sim 11\,\mathrm{km}$ \citep{Capano_2020}) above the pulsar's magnetic poles where the radio beam forms (so-called `polar cap' models) \citep{Caraveo_2014}. However, pulsed optical and gamma-ray emission is possibly emitted at a much greater distance from the pulsar (potentially even outside the light cylinder) \citep{Pétri_2005,hessvela2023}. At least two pulsars are known to produce pulsed gamma-ray emission in the TeV range \citep{magiccrab,hessvela2023}. At a radius at which the confining pressure of the surrounding medium reaches equilibrium with the momentum flux of the wind, a standing termination shock forms \citep{2003MNRAS.345..153L}, where particles can be accelerated further. This termination shock has been directly observed for the Crab pulsar in both the optical and X-rays where it has a `jet-torus' structure (see Figure \ref{fig:struct}) \citep{Hester_2002,mizuno_magnetic_2023}. The wind is dominated by Poynting flux (electromagnetic energy) near its base, but through an unknown mechanism becomes dominated by particle energy by the point of the termination shock. This is known colloquially as the `sigma problem', where $\sigma$ is the ratio of the Poynting flux to particle energy flux in the wind \citep{kirk2003sigmaproblemcrabpulsar,Aharonian:2012zz}. 

One potential solution to the `sigma problem' invokes dissipation of the magnetic field in the wind upstream of the termination shock, a region where magnetic reconnection in the current sheet could also provide an explanation for the flares observed from the Crab nebula \citep{2012AronsCurrentSheet}. Indeed, the equatorial current sheet, a surface separating magnetic field lines in the pulsar wind, is an important location of magnetic field dissipation and reconnection which can naturally lead to efficient particle acceleration \citep{kirk_theory_2009,2011SironiSpitkovsky,2014SironiSpitkovsky}. 

The particles accelerated in the pulsar wind flow out into the surrounding medium forming a Pulsar Wind Nebula (PWN). These regions can be up to tens of parsecs across \citep{Liu_2021}. PWNe produce radiation across the electromagnetic spectrum from the radio up to PeV gamma-rays \citep{706bf6fc-a5ad-35b8-884a-f034d1e2d002,lhaaso}; this review will focus on observations in the $\sim$TeV regime and above. There are two main mechanisms for producing gamma-ray emission in this energy range, Inverse Compton (IC) scattering of background photons by relativistic electrons and positrons, and hadronic production through high-energy p-p interactions that produce $\pi_{\rm 0}$ particles that decay to two photons. Bremsstrahlung may also produce TeV emission due to the deceleration of charged particles in the vicinity of atomic nuclei. Gamma-ray emission from PWNe is largely believed to be dominated by the former, and the multi-wavelength emission from every observed TeV PWN to date can be explained by purely leptonic models (see e.g. \citep{Abeysekara_2018,HESSpwn_population_2018,Dirson_2023}). That said, lepto-hadronic models where there is a mixture of leptonic and hadronic emission have been proposed for a number of sources \citep{Li_2010,Xin_2019,Nie_2022}. The current generation of ground-based gamma-ray instruments, namely the Imaging Atmospheric Cherenkov Telescope (IACT) facilities H.E.S.S. \citep{hess}, MAGIC \citep{magiccrab}, VERITAS \citep{Abeysekara_2018} alongside the particle detector facilities HAWC \citep{2hwc_catalog} and LHAASO \citep{cao2023lhaaso} have put crucial constraints on hadronic acceleration in PWNe. The improved sensitivity, observability and angular resolution of next generation instruments that are planned or being constructed such as CTAO \citep{science_with_CTA}, ASTRI Mini-Array \citep{Scuderi_2022} (IACT facilities) and SWGO \citep{2025SWGOwhitepaper} will provide further insights.

Evidence suggests pulsar wind nebulae evolve over time \citep{HESSpwn_population_2018,Giacinti_2020}. Pulsars are produced with non-negligible proper motions, with 2D values being of the order of $100\,\mathrm{km\,s^{-1}}$ \citep{Hobbs_2005}. Electrons and positrons accelerated at different times and different pulsar positions undergo losses due to IC scattering and Bremsstrahlung, producing an energy-dependent morphology that can vary significantly across wavelengths \citep{hessj18252019,gemingahess2023,Manconi_2024}, with X-ray emission (originating from the highest energy leptons) typically being more compact and closer to the pulsar position than that in the TeV \citep{geminga_caraveo_xray}. Recent results using a joint analysis with H.E.S.S. and the Fermi Large Area Telescope (\emph{Fermi}-LAT) suggests that the Crab Nebula's extension decreases as a function of energy in the GeV-TeV range \citep{Aharonian_2024}. The properties of the environment into which the particles in the PWNe expand, such as the ISM density, can also have a significant effect on the observed PWN morphology \citep{Giacinti_2020}, pre-existing molecular clouds nearby could also provide target material for hadronic gamma-ray emission \citep{Xin_2019}. The H.E.S.S. galactic plane survey showed that the pulsars` offsets from the centroid of their TeV PWNe emission cannot be explained by pulsar proper motion alone \citep{HESSpwn_population_2018}; this is clear evidence for environmental influence on the morphology, such as due to density gradients in the surrounding medium. Typically multiple background photon fields are required to model the IC emission, generally such models consider the cosmic microwave background, emission from dust and background starlight as well as synchrotron photons in the X-ray band produced by the relativistic electrons themselves \citep{atoyan1996,2008ApJ...676.1210Z,Dirson_2023}. At high energies, Klein-Nishina effects are non-negligible \citep{10.1111/j.1365-2966.2005.09494.x}, and can lead to a suppression of the IC emission \citep{Dirson_2023}.

\unskip

At early times, the PWN is fully contained within the parent Supernova Remnant (SNR). The SNR has a forward shock that expands outwards over time, this can produce shell-like gamma-ray emission itself (believed to be largely hadronic in nature) \citep{rxj1713}. The SNR forward shock can also collide with nearby molecular clouds to produce additional gamma-ray emission \citep{ic443_veritas_tev}. Over time, a reverse shock travelling inwards is produced as a result of the forward shock's interaction with the Interstellar Medium (ISM). This reverse shock eventually reaches the PWN, which until this point had been expanding adiabatically \citep{Blondin_2001}, and the PWN is compressed as a result. Competing forces can lead to multiple expansions and contractions of the PWN during this time, a process known as `reverberation'. Modelling this process is complex and so it is not yet completely understood \citep{Bandiera_2023}. After a few tens of kyrs, the pulsar can near the location of the SNR forward shock, leading to pulsar associated particles leaking outside of the parent SNR shell \citep{Giacinti_2020}. Eventually, at times of order 100\,kyr, the pulsar can escape the host SNR completely, though gamma-ray emission can persist due to IC emission from a cloud of relativistic electrons that remain around the pulsar. These so-called `TeV halos' (or pulsar halos) can be extremely large, with TeV emission radii on the sky as large as $3^{\circ}$. Geminga is the canonical example of this type of gamma-ray source, and has been detected by HAWC,  H.E.S.S., LHAASO and Milagro in the TeV regime \citep{geminga_hawc_paper,gemingahess2023,2009ApJMILAGRO}.  

We are now entering an age of Ultra-High-Energy (UHE, $\geq$100\,TeV) gamma-ray astronomy, which began with the detection of the Crab Nebula (the canonical PWN and a significant focus of this review) at energies >100\,TeV in 2019 by the Tibet AS$\gamma$ detector \citep{PhysRevLett.123.051101} and HAWC \citep{Abeysekara_2019}. More than 530 photons have subsequently been detected from the Crab Nebula by LHAASO in the range of 100\,TeV to 1.4\,PeV \citep{lhaaso}. The majority of UHE sources in the first ever catalogue (1LHAASO) in this regime, recently produced by LHAASO, are associated with pulsar systems \citep{cao2023lhaaso}. Whilst the evidence from these observations that such PWNe can be leptonic PeVatrons (sources of PeV cosmic-ray electrons) is reasonably clear, the potential for PeV hadronic cosmic ray production in these sources has not been comprehensively excluded \citep{lhaaso}. Energetic considerations likely rule out the possibility of hadronic processes dominating PWNe gamma-ray emission in the GeV range \citep{atoyan1996}. But the existence of an energetically sub-dominant populations of hadrons in PWNe producing gamma-ray emission at UHE energies has long been speculated \citep{atoyan1996,bp97,Amato_2003}; this scenario has received renewed attention in the light of the recent Crab results \citep{10.1093/mnras/staa2151,lhaaso,Nie_2022,Peng_2022}. Results from 1LHAASO have also challenged the previously established paradigm that the origin of PeV hadronic cosmic rays in our galaxy was associated with SNRs, as the spectra of SNR associated sources typically do not exceed a few tens of TeV \citep{cao2023lhaaso}.

The purpose of this review is to summarise the previous research performed regarding the hadronic population hypothesis, and describe how future observations could constrain this acceleration scenario further. There are essentially three origins for the hadrons that could be accelerated to PeV energies in PWNe; that they are stripped from the pulsar surface and accelerated in the magnetosphere; that hadrons are shocked to PeV energies by the pulsar wind; or that pre-accelerated material is re-accelerated by either the pulsar wind termination shock or the crushing of the PWN by the SNR reverse shock. We will examine each of these scenarios in turn.


\section{Ion escape from the pulsar and acceleration in the wind}
\label{sec:pulsarorigin}

\subsection{Ions originating from the pulsar}
\label{sec:psrions}

\citet{cheng} were the first to propose that protons could be stripped from the surface of the Crab pulsar and accelerated in the pulsar magnetosphere to $\sim$\,PeV energies. They proposed that overdense electrons are accelerated in the `outer gap', a small region within the pulsar magnetosphere with a finite electric potential \citep{10.1093/mnras/stu2564}. In such gaps the electric field has a component aligned with the magnetic field; this results in the conversion of the rotational energy of the pulsar into particle kinetic energy. Various gap models at different locations for particle acceleration in the pulsar magnetosphere have been proposed \citep{1983ApJ...266..215A,1986ApJ...300..500C}. \citet{cheng} propose the accelerated electrons then move inwards to the pulsar and strip protons from its surface, after which the protons too are accelerated in the outer gap. In this model $\sim$\,PeV protons escape the pulsar magnetosphere, with the condition that $\vec{\Omega}\cdot\vec{\mu}>0$ where $\vec{\mu}$ is the magnetic moment of the pulsar. These protons then interact with gas in the nebula to produce TeV gamma-ray emission in their model, assuming a high gas density of $n_{\rm H}=500\,\mathrm{cm^{-3}}$.

\citet{bednarek2003} also predict the production of $\sim$PeV protons in the Crab Nebula, albeit with a different origin. They propose that it is in fact iron nuclei that are stripped from the pulsar surface, which then photodissociate releasing protons and neutrons, the latter of which eventually decay to protons, electrons and electron anti-neutrinos. They then propose that these protons interact with the material in the nebula, considering that some neutrons may not be trapped within the nebula before they decay. Specifically they argue that the gamma-ray flux could be enhanced if the resultant protons are trapped within filamentary structures (with density $n_{\rm H}=500\,\mathrm{cm^{-3}}$) within the nebula, as previously suggested by \citet{aharonian_atoyan}. These filaments are likely formed as a result of Rayleigh-Taylor instabilities at the boundary between the PWN and the surrounding supernova ejecta, and can be observed in optical and IR measurements of the Crab (see Figure \ref{fig:struct}) \citep{1997ApJS..109..473B,Loh_2010}. Magnetohydrodynamical simulations support the formation of high density filaments \citep{2014PorthRTcrab}. However it should be noted that given the Crab Nebula's mass (gas plus dust) of $7.2\pm0.5\,\mathrm{M_{\odot}}$ and its radius of 2.1\,pc the average gas density in the nebula $n_{\rm H}$ is likely of order $\sim5\,\mathrm{cm^{-3}}$ \citep{10.1093/mnras/staa2151}. Particle-in-cell (PIC) simulations of pulsar magnetospheres also predict the scenario of ions being stripped from the pulsar surface, but only in the $\sim$MeV energy range in the specific case of an aligned rotator (i.e. where the pulsar magnetic axis and rotation axis are aligned) \citep{Chen_2014}; however it should be noted that the conclusions of PIC simulation studies are non-trivial to extrapolate to astrophysical scales \citep{Lemoine_2015}.

A key quantity that is an indicator of whether protons or ions are escaping the pulsar surface into the pulsar wind is the average pair-production multiplicity $\langle\kappa\rangle$. As electrons and positrons escaping the pulsar surface undergo gamma-ray bremsstrahlung and then pair production in the pulsar magnetosphere producing particle cascades, $\langle\kappa\rangle$ describes the number of $e^+e^-$ pairs escaping the light cylinder per electron (or positron) stripped from the pulsar surface. It is given by \citep{de_Jager_2007}
\begin{equation}
\langle \kappa \rangle = \frac{N_{\rm el}}{2N_{\rm GJ}}\,,
\label{eq:kappa}
\end{equation}
where $N_{\rm el}$ is the number of electrons in the PWNe and $N_{\rm GJ}$ is the so-called Goldreich-Julian density given by \citep{1969ApJ...157..869G}
\begin{equation}
N_{GJ}=\int_{t=0}^{t=-\tau(P_0)} \frac{[6c\dot{E}(t)]^{1/2}}{e} (-dt)\,.
\label{eq:gj}
\end{equation}
Here, $\tau$ is the pulsar age as a function of pulsar birth period $P_0$. Due to the lower Bremsstrahlung cross-section, protons and other ions do not multiply in such cascades \citep{kirk_theory_2009}, so can only make up $1/{\langle\kappa\rangle}$ of the total particles in the PWN at most. Although there is no model-independent means of estimating $\langle\kappa\rangle$ for any given pulsar, two studies have provided lower-limits for a small number of pulsars associated with PWNe that have been observed in the TeV \citep{de_Jager_2007,Spencer_2025}.

A number of works consider the scenario that pulsars with millisecond birth periods \citep{2000BlasiUHECRpulsar,kefang_2013,2015JCAP...08..026K,Lemoine_2015} can accelerate ions to very-high-energies ($\sim 10^{20}\,\mathrm{eV}$), either as they are stripped from the pulsar surface by its electric field \citep{2000BlasiUHECRpulsar,2015JCAP...08..026K} or in the wind \citep{Lemoine_2015}. Related studies have suggested that an extragalactic population of pulsars with such properties could reproduce the observed UHECR flux \citep{kefang_2013} by assuming a Gaussian distribution of pulsar birth periods centered on 300\,ms and a sufficiently large volume. This model is based on the work \citep{2006ApJ...643..332F} (also used as a basis for \citet{2015JCAP...08..026K}), however \citet{2006ApJ...643..332F} state that the pulsar birth period distribution is poorly constrained by their model, and so is not necessarily evidence for the existence of $\sim$ms birth period pulsars. \citet{2003ChJAA...3..166H} postulated a theoretical model for predicting the birth periods of pulsars based on their observed proper motions that resulted in birth periods in the range $0.6-2.6$\,ms, however this is now likely disproven as it would over-predict the number of observed gamma-ray PWNe \citep{2008ApJ...678L.113D}. An alternative motivation for ms birth period pulsars was an analysis of the radio spectrum of the Crab Nebula by \citet{atoyan1999radiospectrumcrabnebula}, which suggested the pulsar might have a $\sim$5\,ms birth period. However, more recent modelling has suggested it is rather $\sim18.3\,\mathrm{ms}$ \citep{Zhang_2022} or $\sim$15.6\,ms \citep{araujo2023modellingbrakingindexisolated,10.3389/fspas.2024.1390597}. As such, we will consider ms birth period hadronic acceleration scenarios as being unlikely, for the particular case of galactic sources that we consider in this review. \citet{guepin2020} also postulate that pulsars with millisecond periods (without the need for this to be the case at birth) could produce >PeV protons through magnetic reconnection in the magnetosphere. However millisecond period pulsars spun-up through accretion are expected to be sufficiently old such that a >TeV gamma-ray PWN powered by electrons would have since faded away.

\subsection{Ions in the pulsar wind}
\label{sec:massloading}

Mass-loading of pulsar winds by nuclei has been explored in various studies \citep{2003LyutikovLoading,2015MorlinoBSPWNloading,2019KirkGiacinti}, demonstrating that this channel is entirely feasible with current theory, yet remains disfavoured as a dominant component to the majority of pulsar winds studied experimentally. Neutral hydrogen atoms from the ISM may be picked-up by and ionized in the pulsar wind, subsequently accelerated as part of the plasma flow. Such mass-loading (beyond the pulsar wind termination shock) increases the particle energy flux and decreases the Poynting flux, affecting the pulsar wind flow dynamics and leading somewhat counter-intuitively to a net increase in flow velocity \citep{2003LyutikovLoading}. Whilst this scenario has been explored for both ram pressure confined and bow shock PWNe, well-matching H$\alpha$ and X-ray observations, the ion density within the pulsar wind remains small \citep{2003LyutikovLoading,2015MorlinoBSPWNloading}. Nevertheless, mass-loading scenarios provide a potential solution to the `$\sigma$-problem' of pulsar winds \citep{2012AronsCurrentSheet}. \citet{2019KirkGiacinti} show that the presence of ions in relativistic flows dominated by Poynting-flux can lead to inductive acceleration, raising the energy at which electrons are injected into the pulsar wind. In this case, the maximum energy achieved approaches the Hillas limit and comparable powers occur in both leptonic and ion components.  

The so-called `resonant cyclotron absorption' model has also been postulated as a means of accelerating ions in the pulsar wind \citep{hoshino,amatoarons}. In this, ions are assumed to already be present in the pulsar wind. Positrons and electrons are accelerated as a result of the absorption of resonant frequency waves emitted; such models can reproduce the optical to X-ray emission of the Crab, however they cannot reproduce the radio to IR emission \citep{kirk_theory_2009}. The ions themselves are also accelerated in this scenario \citep{Amato_2003}, but for this to be able to account for PeV emission (at least in the case of the Crab), a Lorentz factor of the wind of the order $10^8-10^9$ would be required. Whilst not exactly known, other studies place the Lorentz factor in the range $10^4-10^7$ for the Crab \citep{kirk_theory_2009,cerutti2020,amato2021}. 

\section{Hadrons originating external to the PWN}
\subsection{Shock Mixing from the ambient environment}
\label{sec:mixing}

Particles escaping from the pulsar surface and/or being re-accelerated in the wind is not the only potential means of producing TeV-PeV hadrons in a PWNe. According to a mechanism proposed by \citet{bell_cosmic_1992}, a cosmic ray from a supernova remnant may, depending on its trajectory, be able to enter a PWN and experience acceleration due to the magnetic field. Through this process it could gain a factor $\sim100-1000$ in energy \citep{1994MNRAS_LucekBell}. Some evidence for such mixing is garnered via the presence of Rayleigh-Taylor instabilities in the Crab nebula, suggesting that material from the under-luminous confining remnant is able to penetrate within the nebula \citep{1984KennelCoronitiCrab}. Magnetohydrodynamical simulations of the Crab nebula have since confirmed that the presence of these instabilities does indeed lead to flow-mixing \citep{2014PorthRTcrab}. The presence of nuclei within these filaments is observationally well-established (e.g. \citep{1990HudginsNiFeCrab}), with the most recent measurements of Ni/Fe abundance ratios coming from the James Webb Space Telescope (JWST), (see \citet{2024TemimJWSTcrab} and references therein). 

Two recent works have considered the possibility that particles accelerated during the initial supernova blast wave are then re-accelerated, either by reverse shock interactions compressing the PWN as suggested by \citet{Ohira_2018}, or through them diffusing back across the PWN boundary and being re-accelerated at the pulsar wind termination shock \citep{1994MNRAS_LucekBell,Spencer_2025b}. The attraction of these models is that they can physically explain the mechanism by which hadrons are accelerated to PeV energies and produce gamma-ray emission without having to introduce additional hadronic populations by-hand, or without having to make strong assumptions about the nature of the wind. \citet{Ohira_2018} do however neglect the effect of adiabatic losses in their model.  


\subsection{Hadronic population of unknown origin in the PWN}

A number of works introduce proton populations in the PWNe of the Crab in order to reproduce the UHE results. \citet{Peng_2022} introduce a power law with exponential cutoff spectrum of protons as accelerated at the TS, referring to \citet{1984ApJ...283..710K}, who do not discuss the magneto-hydrodynamic mechanism for the acceleration of ions. However, \citet{Peng_2022} do successfully fit a complex spatially-dependent lepto-hadronic model to the Crab spectrum. 
\citet{10.1093/mnras/staa2151} had previously used a similar approach for the HAWC and Tibet AS$\gamma$ data albeit with a much simpler model, and constrained the pulsar luminosity conversion efficiency factor into hadrons to be $\sim0.5$\%. \citet{Liu_2021} perform a similar study, yet accounting for the effects of protons escaping the nebula, find the luminosity conversion efficiency factor for protons to be up to 50\%.  \citet{Nie_2022} similarly refer to \citet{cheng} and \citet{bednarek2003} as motivation for their lepto-hadronic model, but were required to introduce an additional proton spectrum of unknown origin with a power law with exponential cutoff spectrum.

The discovery of VHE gamma-ray emission from the nearby PWN Vela\,X was remarkable for its pronounced spectral maximum at around 1\,TeV, which was well-modelled under a leptonic emission scenario \citep{2006VelaXdiscovery}. Only a comparatively low efficiency of energy conversion from spin-down power into energetic particles was required $\sim10^{-3}$, which led \citet{2006HornsVelaX} to suggest that a more reasonable fraction of the spin-down power could be accounted for in particles if the gamma-ray emission was hadronic in origin. Interestingly, they find that the gamma-ray SED can be accounted for by a proton population for even modest ambient densities of $0.6\,\textrm{cm}^{-3}$ consistent with observations, or alternatively by iron (or lighter) nuclei loaded in the pulsar wind. Along similar lines, \citet{2009ZhangVelaXleptohadronic} find that a lepto-hadronic model, with iron nuclei accounting for the TeV emission and an electron population accounting for the X-ray emission from the nebula, is compatible with the observational data on Vela\,X. In later works, however, the MWL SED from Vela\,X has been almost exclusively modelled as originating from an energetic population of electrons \citep{velax_escape,velax_tibaldo,VelaXslowdiff}. 

In a study of the complex system VER\,J2229+608, comprised of both the SNR G106.3+2.7 and a PWN powered by the energetic pulsar PSR\,J2229+6114, \citet{Xin_2019} suggest that the high energy component could be powered by hadrons from the PWN rather than the SNR, given the source morphology and low energy content of CRs in the SNR. 
Indeed, \citet{2024ApJChenBoomerang} also find that inverse Compton emission alone is likely insufficient to account for all of the gamma-ray emission from this source, suggesting that a fraction of the flux could originate from a hadronic particle population. 

More generally, lepto-hadronic scenarios such as these remain comparatively infrequent in modelling of PWNe and require further detailed observations to disentangle multiple components in such composite systems. 

\section{Observational constraints on the presence of hadrons in PWNe}

Given that PWNe are leptonically dominated systems, there are a limited number of ways in which the presence of hadrons in PWNe can be constrained observationally. These fall into essentially three categories: gamma-ray spectral signatures; gamma-ray emission spatially associated with target gas in the vicinity of a PWN; and the detection of neutrinos in coincidence with PWNe. The latter is the only direct channel that can unequivocally be attributed to the presence of hadrons. 

\subsection{Neutrinos in coincidence with PWNe}
To date, there have been no detections of neutrinos associated with PWNe, however constraints have been placed via stacking analyses. A study performed by the IceCube collaboration using 9.5 years of data from the IceCube neutrino observatory found no significant excess based on a stacking analysis of 35 TeV-bright PWNe \citep{Aartsen_2020}. Several different weighting schemes were tested, including according to gamma-ray flux, pulsar spin frequency and pulsar characteristic age, although the largest excess of 40 events was found under equal weighting, and remained insignificant compared to the isotropic background. 
Their results hence showed that PWNe cannot be hadronically dominated \citep{Aartsen_2020}.
In the context of the detection of neutrino emission from the galactic plane \citep{2023SciIceCube_galplane}, additional searches were performed using a stacking analysis of catalogued gamma-ray sources. The pre-trial significance for the 12 PWNe tested was $3.24\,\sigma$, equal to that for SNRs and comparable to the estimate for catalogued sources of unidentified nature \citep{2023SciIceCube_galplane}. Although an upper limit was placed on the neutrino flux from these 12 PWNe, the results to date do not yet offer meaningful constraints on the hadronic composition of PWNe compared to other galactic accelerators. 

\citet{2024ApJ.Gagliardini} estimated the expected neutrino flux that could be detectable by KM3NeT (a neutrino facility under construction) from single sources provided that all of the observed TeV emission is hadronic in nature. Under these assumptions, the brightest PWNe, such as Vela\,X would be detectable by KM3NeT at $\sim7\,\sigma$ within one year. A lack of significant detection can hence be taken as an indication that the associated emission is not fully hadronic in nature. \citet{2017ApJ.DiPalma} estimated the neutrino flux from individual PWNe following initial constraints from IceCube results. This included the expectations for individual sources that could be detected by ANTARES (a now decommissioned neutrino detector) or KM3NeT, with Vela\,X and MSH\,15-52 identified as the most promising targets, whilst the non-detection of the Crab by IceCube was used to derive a constraint on the Crab nebula of $\leq15\%$ hadron content in the PWN \citep{2017ApJ.DiPalma}. 

Studies investigating individual source flux estimates or stacking analyses are inherently limited to the size of the catalogue used to define the sample, therefore intrinsically biasing resulting constraints on the true underlying population. In particular, due to the pulsar beaming fraction several unidentified gamma-ray sources may additionally be related to pulsars. To address this, \citet{liang2025potentialcontributionyoungpulsar} construct a synthetic population of young PWNe, finding that under an optimistic scenario young PWNe could contribute up to 5\% of the total neutrino flux measured by IceCube at 100\,TeV. 

In future, improved constraints across a broad population of PWNe can be obtained by combined analyses, such as the combination of both track-like and cascade-like events in IceCube data, or of neutrino datasets jointly with other instruments. The sensitivity of such combined event analyses is set to outperform current analyses, due to the broader sky coverage that can be afforded \citep{2025ApJ_IceCube_trackcascade}. 
Joint analyses of datasets from multiple instruments enable in-depth studies yielding deeper insights than otherwise possible. \citet{2024EPJC_UnbehaunJoint} demonstrated the potential for future joint analyses of KM3NeT and CTAO data; combining a 10\,year exposure of KM3NeT with 200\,hours of CTAO observations could constrain the hadronic fraction present in the PWN Vela\,X to $\leq15\%$. 

\subsection{Gamma-ray spatial and spectral signatures}

The spatial coincidence of gamma-ray emission with molecular clouds or gas can provide a strong indicator for the presence of hadrons. Such interstellar material acts as a target for proton-proton interactions, producing gamma-ray emission due to the decay of neutral pions. This can lead to a local enhancement of gamma-ray emission, even in the case of a dominantly leptonic parent particle population. However, excellent angular resolution and careful modelling are required to identify such cases. 

An example of such a system is the gamma-ray bright PWN HESS\,J1825-137, which is located in the vicinity of multiple molecular clouds \citep{2016MNRAS.VoisinTurbulence}. During the early stages of SNR expansion, it is thought that the SNR forward shock interacted with a molecular cloud towards the north of the pulsar PSR\,J1826-1334, leading to the early formation of a reverse shock that consequently swept the nebula away, causing its preferential expansion in a southerly direction. Enhanced turbulence observed in the molecular cloud located towards the north of the pulsar lends support to this idea \citep{2016MNRAS.VoisinTurbulence}. 
In a detailed analysis by the HAWC collaboration, the gamma-ray emission in this region was modelled with three gaussian components, two of which approximately corresponded to HESS\,J1825-137 and HESS\,J1826-130 respectively \citep{2021HAWCJ1825}. The third residual component was tentatively associated to a molecular cloud in the region, although the analysis is also limited by the angular resolution of the facility and H.E.S.S. observations of the same region do not clearly support this interpretation \citep{hessj18252019}. Such gamma-ray emission coincident with a cloud could indicate the presence of hadrons in the vicinity of this PWN, however it would  remain unclear if the hadrons originate from the PWN itself or from its progenitor SNR. \citet{voisin2017} performed modelling studies of this region, demonstrating that whether hadronic particles originate from the SNR or the PWN, the resulting gamma-ray emission map will appear nearly identical; localised to the molecular cloud. 
This highlights the key difficulty with using coincident target material to identify the presence of hadrons; which is that the original accelerator of that hadron population remains nevertheless unconstrained. 

\begin{figure}
    \centering
    \includegraphics[width=0.7\linewidth]{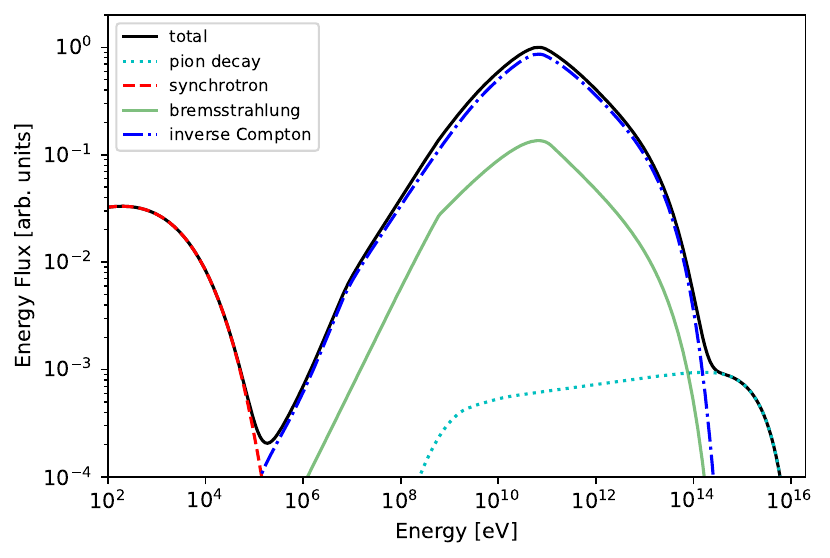}
    \caption{Example spectral energy distribution for a Crab-like PWN, with arbitrary flux normalisation. A potential hadronic contribution could emerge at the highest energies, beyond the Klein-Nishina cut-off. }
    \label{fig:leptohadronicSED}
\end{figure}

A plausible alternative is to search for signature pion-decay bump features in the spectral energy distribution of PWNe towards the highest energies. The inverse Compton scattering of electrons proceeds in two regimes, the Thomson regime at lower energies and the Klein-Nishina regime at high energies, in which the electron loses a substantial fraction of its energy on each scattering interaction \citep{BlumenthalGould1970}. Due to these severe cooling losses, leptonic spectra may exhibit a cut-off at energies $\gtrsim100$\,TeV, such that emission beyond these energies has been invoked as potential evidence of a hadronic component \citep{2004vhec.bookAharonian,Nie_2022}. 

Figure \ref{fig:leptohadronicSED} illustrates a potential lepto-hadronic scenario, based on modelling of the Crab nebula by \citet{Nie_2022}, in which the proton particle population has 10\% of the total energy in the accelerated particle population. A broken power law spectrum with an exponential cut-off is assumed for the electrons and power law with exponential cut-off for the protons, with the modelling code GAMERA used \citep{gamera}. 
However, such scenarios typically rely on a good characterisation of the spectral energy distribution towards the high-energy end of the spectrum, a region subject to poor statistics and close to the sensitivity threshold of most instruments. 

\citet{2021ApJ_Breuhaus} demonstrated that particularly in high radiation environments, IC scattering may continue to produce radiation beyond 100\,TeV. To date, even in light of spectra from multiple PWNe reaching energies $\gtrsim100$\,TeV and in the case of the Crab nebula reaching $\gtrsim1$\,PeV \citep{lhaaso}, there is no conclusive evidence for such a hadronic component to the spectrum emerging at the highest energies. 

\citet{Liu_2021} alternatively examined the constraints that current observational data on the Crab nebula place on the proton content of the pulsar wind, finding that as much as 10-50\% of the pulsar spin-down energy could be converted to protons and remain consistent with observations. 


In light of multiple modelling studies concerning the potential for a hadronic contribution to the Crab nebula spectrum emerging at the highest energies, it is intriguing to note the presence of a weak hot spot within the IceCube sky map at approximately the location of the Crab nebula \citep{2023SciIceCube_galplane}. However, as this region is not yet statistically significant ($\lesssim 3\,\sigma$), such tentative associations remain speculative awaiting further observations. 

\subsection{Ultra High Energy Cosmic Rays}

As speculated in section \ref{sec:pulsarorigin}, pulsars with millisecond birth periods and magnetars have been hypothesised as potential sources of ultra-high-energy (UHE, $E\gtrsim 10^{18}$\,eV) CRs \citep{2003AronsMagnetar}. At these energies, charged CRs will be less significantly deflected than at lower energies, such that their arrival direction may also indicate their direction of origin. Also, due to CR interactions with the CMB and thereby losing energy (the GZK effect), there is a limited horizon from within which such UHECRs may originate, typically $\lesssim200$\,Mpc. If pulsars with ms birth periods and/or magnetars are a significant source of UHECRs, an excess of CR events would be expected towards either the plane of the Milky Way, or towards nearby galaxies. To date, there are mild indications for anisotropies in arrival directions of UHECRs at the $\sim4\sigma$ level, yet no excess that distinctly correlates with the plane of the Milky Way \citep{2022ApJ.PAO_arrivaldirns,2018ApJ.PAOanisotropy}. A galactic UHECR point source with a relatively light cosmic ray composition has also likely now been excluded by the Auger detector \citep{2022ApJ.PAO_arrivaldirns}. Given the near isotropic distribution of UHECRs, a lower bound was estimated for the density of uniformly distributed sources of $(0.06-5)\times10^{-4}$\,Mpc$^{-3}$, already implying an origin outside of the our own galaxy. Indeed, searches for candidate sources are generally conducted using lists of other galaxies \citep{2018ApJ.PAOanisotropy,2022ApJ.PAO_arrivaldirns}. 

\section{Implications for the galactic cosmic ray flux}
\label{sec:CRflux}

In the case that hadrons from the progenitor SNR or ambient medium are re-accelerated by pulsar wind nebulae, the observed gamma-ray spectrum from individual sources may harden towards the highest energies. Such a mechanism can provide a small boost to the source flux beyond $\sim$100\,TeV, but is unlikely to contribute markedly to the overall galactic cosmic ray flux. The total energetics across a Milky Way source population (e.g. of SNRs) remain unchanged by considering such scenarios, with re-acceleration only relevant in considering the specifics of individual sources. 

Irrespective of the acceleration mechanism, the maximum energy achievable within pulsar environments is limited by the potential drop, available in the electric field, which directly relates to the pulsar spin-down energy. In order to achieve energies around a PeV, a spin-down luminosity $\dot{E}\gtrsim 10^{36}$\,erg/s is required \citep{deonaWilhelmi2022,amato2021}. This essentially prohibits a contribution from middle-aged pulsars to hadronic CRs at PeV energies. 

\begin{figure}
    \centering
    \includegraphics[width=0.7\linewidth]{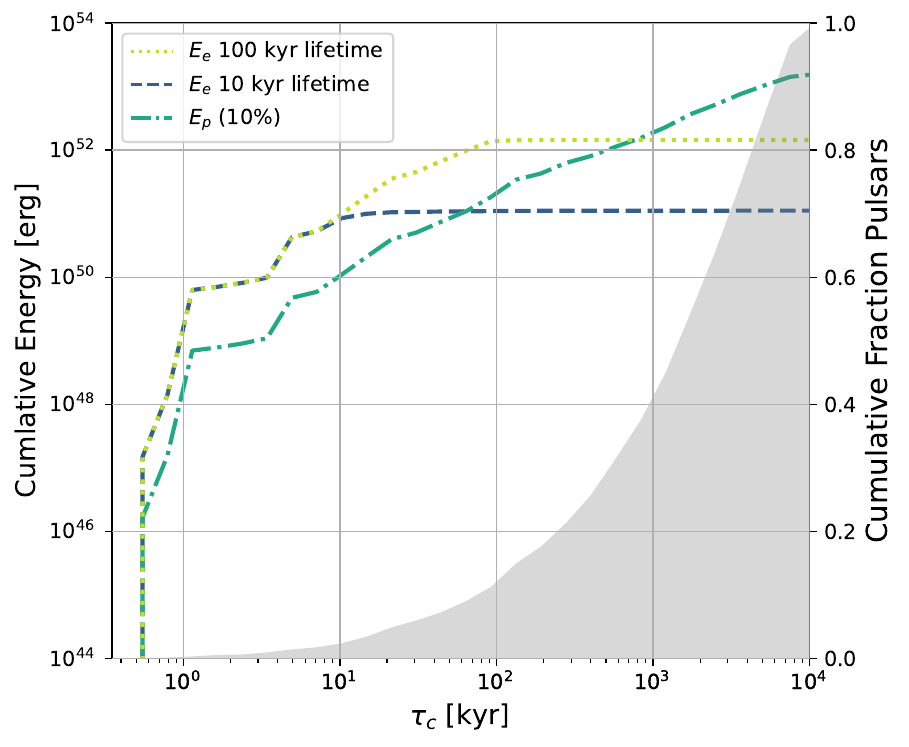}
    \caption{Cumulative energy in electrons and protons originating from pulsars in the Milky Way, assuming that 10\% of the energy in particles goes into protons. }
    \label{fig:Epfrac_pwne}
\end{figure}

In the case that hadrons are released as part of the pulsar wind, directly originating from the neutron star surface, this may represent an additional contribution to the galactic cosmic ray flux not generally considered. To estimate the influence of this component, we estimate the energetics of the contribution by following the argument of \citet{Giacinti_2020}, yet adapted for hadrons. 
Figure \ref{fig:Epfrac_pwne} shows the cumulative energy output from galactic pulsars as listed in the ATNF \citep{manchester_atnf_2005} (and excluding millisecond pulsars) as a function of their characteristic age, compared to the cumulative fraction of pulsars in the sample. The energy output is estimated from the beaming fraction corrected pulsar spin-down luminosity, and is shown for electrons integrated over $10^4$\,yr and $10^5$\,yr timescales, corresponding to the radiative lifetime of $\sim10$\,TeV and $\sim1$\,TeV electrons in the ISM respectively. 
The pulsar energy output is assumed to evolve as $\left(1+\frac{t}{\tau_0}\right)^{-\alpha}$ with spin-down timescale $\tau_0=10^3$\,yr and $\alpha=(n+1)/(n-1)=2$ for a pulsar braking index of $n=3$. 
Additionally, we assume here that 90\% of the energy output goes into electrons and 10\% into protons, where protons are taken to represent all hadrons present (as they are likely the majority species \citep{2015JCAP...08..026K}). 

One can immediately see that whilst the cumulative energy in electrons with different lifetimes saturates, the proton contribution rises corresponding to the pulsar population, as the fraction remains constant and protons have a much longer lifetime in the ISM. Nevertheless, the total energy carried by protons in this scenario peaks at $\sim10^{53}$\,erg and reaches $\sim10^{48}$\,erg for 1\,kyr, (i.e. $\sim10^{38}$\,erg/s) which is just a few percent of the amount required to account for the total power in CRs in our galaxy of $\sim10^{41}$\,erg/s \citep{2008Hoerandel}. 
As the assumed spin-down timescale $\tau_0$ is a highly uncertain parameter, we estimate its influence on our results by varying $\tau_0$ between 100\,yr and 10\,kyr, finding that the cumulative energy remains within a factor $\sim2-3$ (at 1\,kyr) and up to a factor 10 at most (at 10\,kyr) of the curves shown in Figure \ref{fig:Epfrac_pwne}.
Even in optimistic scenarios in which protons could carry as much as $\sim30$\% of the pulsar energy output, this is likely to only be feasible for a few specific pulsars and not on average for the population, and would be insufficient to significantly contribute to the galactic CR flux. 

\begin{table}[H]
    \caption{Summary comparing the key aspects of the primary scenarios for hadronic particle acceleration in pulsar environments. }
    \label{tab:summary}
        \setlength{\tabcolsep}{3pt}
  \begin{adjustwidth}{-\extralength}{0cm}
\begin{minipage}{\fulllength}
\setlength{\cellWidtha}{\textwidth/4-2\tabcolsep-0.05in}
\setlength{\cellWidthb}{\textwidth/4-2\tabcolsep-0.05in}
\setlength{\cellWidthc}{\textwidth/4-2\tabcolsep-0in}
\setlength{\cellWidthd}{\textwidth/4-2\tabcolsep+.1in}
\begin{tabularx}{\textwidth}{>{\centering\arraybackslash}m{\cellWidtha}>{\centering\arraybackslash}m{\cellWidthb}>{\centering\arraybackslash}m{\cellWidthc}>{\centering\arraybackslash}m{\cellWidthd}}
\toprule

    Scenario & Ions from the Pulsar & Mass-loading of the Wind & Shock Mixing / Re-acceleration \\
    \midrule
    Physical origin of hadrons & extracted from NS surface & ISM & SNR / reservoir \\ \midrule
    Main acceleration site & vacuum gap in pulsar magnetosphere & pulsar wind (beyond TS) & PWN + SNR reverse shock interaction or at PWN TS \\ \midrule
    Achievable energies & $\sim 10^{20}$\,eV (ms birth periods) or $\sim10^{15}$\,eV (if ms now) & approaching Hillas limit, potentially PeV for a Crab-like system & a factor $\sim100-1000$ energy increase, e.g. into TeV-PeV range\\
    \midrule
    Notes & ms birth period required for energies $\sim10^{20}$\,eV, yet highly unlikely. Low $\langle\kappa\rangle$ required. & Density of surrounding ISM can be low, process suggested as a solution to the $\sigma$-problem. & Re-injection and re-acceleration only feasible at early times, requires Bohm diffusion to reach the TS and PeV energies. \\ \midrule
    Section & \ref{sec:psrions} & \ref{sec:massloading} & \ref{sec:mixing} \\
    \bottomrule
    \end{tabularx}
\end{minipage}
\end{adjustwidth}
\end{table}

\section{Outlook}

The presence of hadrons in pulsar wind nebulae is compatible with our current theoretical understanding, with essentially three different potential origins for this ionic component: nuclei stripped from the surface of pulsars with ms periods; mass-loading of the pulsar wind from the surrounding medium; and shock mixing due to interactions with the progenitor supernova remnant. A summary of the key properties of these three scenarios is provided in table \ref{tab:summary}.
The prospect of PWNe harbouring an energetically significant hadronic particle population that contributes to the galactic CR flux is an exciting one, given the preponderance of pulsar-associated sources in the galactic VHE gamma-ray source population. However, first estimates indicate that the contributions of hadrons in pulsar environments are likely to be sub-dominant, both for individual sources and on the scale of the whole Milky Way source population. As outlined in section \ref{sec:CRflux}, the estimated total energetic contribution of protons from PWNe to the total CR flux is far lower than that necessary to sustain the CR power in our galaxy. 

Observational channels to constrain this scenario are however limited, as the predicted spectral and spatial signatures are second-order compared to the dominant non-thermal emission from PWNe. Searches for potential pion-bump signatures in gamma rays rely on high photon statistics and good energy resolution at energies $\gtrsim100$\,TeV, whilst resolving gamma-ray or neutrino emission coincident with target gas material relies on excellent angular resolution. 
These requirements pose a challenge to both current and next-generation gamma-ray and neutrino detectors, for which the projected flux sensitivity and resolution is unlikely to be sufficient to probe this scenario in individual sources. 
One alternative is to approach the problem statistically, placing constraints on PWNe as a population e.g. via stacking analyses using neutrino datasets. 
Further broad-band multi-wavelength studies of PWNe are required to exclude high-energy leptonic emission in most cases, given the possibility of secondary electron fields. 
To date, the Crab Nebula is the only significantly studied target across the electromagnetic spectrum \citep{Dirson_2023}, with strong evidence for the presence of nuclei in filaments that extend into the nebula, providing some support for shock-mixing scenarios \citep{2024TemimJWSTcrab}. Simulations indicate that mass-loading of pulsar winds from the surrounding medium may provide a solution to the `$\sigma$-problem' whilst also reproducing H$\alpha$ and X-ray features seen around some PWNe. 

Further detailed studies are necessary, combining observational data with theory to better constrain the presence of hadrons in PWNe. With multiple channels available, recent results have demonstrated that hadronic particle acceleration in PWNe remains not only a viable but also a highly plausible scenario, that may well be realised in nature. 

\funding{This work was supported by the Deutsche Forschungsgemeinschaft (DFG) project number 452934793.}

\acknowledgments{The authors thank the anonymous referee for valuable comments that helped to improve the manuscript. }

\conflictsofinterest{The authors declare no conflicts of interest.}

\reftitle{References}

\bibliography{references.bib}

\end{document}

%% file: journalcommands.tex
\newcommand*\aap{A\&A}
\let\astap=\aap
\newcommand*\aapr{A\&A~Rev.}
\newcommand*\aaps{A\&AS}
\newcommand*\actaa{Acta Astron.}
\newcommand*\aj{AJ}
\newcommand*\ao{Appl.~Opt.}
\let\applopt\ao
\newcommand*\apj{ApJ}
\newcommand*\apjl{ApJ}
\let\apjlett\apjl
\newcommand*\apjs{ApJS}
\let\apjsupp\apjs
\newcommand*\aplett{Astrophys.~Lett.}
\newcommand*\apspr{Astrophys.~Space~Phys.~Res.}
\newcommand*\apss{Ap\&SS}
\newcommand*\araa{ARA\&A}
\newcommand*\azh{AZh}
\newcommand*\baas{BAAS}
\newcommand*\bac{Bull. astr. Inst. Czechosl.}
\newcommand*\bain{Bull.~Astron.~Inst.~Netherlands}
\newcommand*\caa{Chinese Astron. Astrophys.}
\newcommand*\cjaa{Chinese J. Astron. Astrophys.}
\newcommand*\fcp{Fund.~Cosmic~Phys.}
\newcommand*\gca{Geochim.~Cosmochim.~Acta}
\newcommand*\grl{Geophys.~Res.~Lett.}
\newcommand*\iaucirc{IAU~Circ.}
\newcommand*\icarus{Icarus}
\newcommand*\jcap{J. Cosmology Astropart. Phys.}
\newcommand*\jcp{J.~Chem.~Phys.}
\newcommand*\jgr{J.~Geophys.~Res.}
\newcommand*\jqsrt{J.~Quant.~Spectr.~Rad.~Transf.}
\newcommand*\jrasc{JRASC}
\newcommand*\memras{MmRAS}
\newcommand*\memsai{Mem.~Soc.~Astron.~Italiana}
\newcommand*\mnras{MNRAS}
\newcommand*\na{New A}
\newcommand*\nar{New A Rev.}
\newcommand*\nat{Nature}
\newcommand*\nphysa{Nucl.~Phys.~A}
\newcommand*\pasa{PASA}
\newcommand*\pasj{PASJ}
\newcommand*\pasp{PASP}
\newcommand*\physrep{Phys.~Rep.}
\newcommand*\physscr{Phys.~Scr}
\newcommand*\planss{Planet.~Space~Sci.}
\newcommand*\pra{Phys.~Rev.~A}
\newcommand*\prb{Phys.~Rev.~B}
\newcommand*\prc{Phys.~Rev.~C}
\newcommand*\prd{Phys.~Rev.~D}
\newcommand*\pre{Phys.~Rev.~E}
\newcommand*\prl{Phys.~Rev.~Lett.}
\newcommand*\procspie{Proc.~SPIE}
\newcommand*\qjras{QJRAS}
\newcommand*\rmxaa{Rev. Mexicana Astron. Astrofis.}
\newcommand*\skytel{S\&T}
\newcommand*\solphys{Sol.~Phys.}
\newcommand*\sovast{Soviet~Ast.}
\newcommand*\ssr{Space~Sci.~Rev.}
\newcommand*\zap{ZAp}